\documentclass{JHEP3}

\usepackage{epsfig}
\usepackage{amsmath}
\usepackage{amssymb}
\def\e{\,{\rm e}\,}
\def\d{{\rm d}}
\def\i{{\rm i}}
\def\D{{\cal D}}
\def\ym{y_{\rm min}}
\def\del{\varepsilon}
\def\eps{\varepsilon}
\def\x{\sigma}
\def\om{\omega}
\def\omb{\bar{\omega}}
\def\tr{{\rm tr}\,}
\newcommand{\rf}[1]{(\ref{#1})}
\newcommand{\eq}[1]{Eq.~(\ref{#1})}
\def\be{\begin{equation}}
\def\ee{\end{equation}}
\def\bea{\begin{eqnarray}}
\def\eea{\end{eqnarray}}
\def\LA{\left\langle}
\def\RA{\right\rangle}
\newcommand{\non}{\nonumber \\*}

\preprint{ITEP--TH--28/12}

\title{More about One-Loop Effective Action of Open\\ Superstring in 
$\boldsymbol{AdS_5\times S^5}$} 

\author{Charlotte Kristjansen \\
The Niels Bohr Institute,\\
\mbox{Blegdamsvej 17, 2100 Copenhagen \O, Denmark}\\
E-mail: \email{kristjan@nbi.dk} }

\author{Yuri Makeenko\\ 
Institute of Theoretical and Experimental Physics,\\
B.~Cheremushkinskaya 25, 117218 Moscow, Russia \\
E-mail: \email{makeenko@itep.ru}
}

\date{June 25, 2012}

\abstract{
We reconsider the calculation of the one-loop effective action for an open
Green--Schwarz superstring in the $AdS_5\times S^5$ background
for a circular boundary loop. By an explicit computation of the ratio
of relevant determinants, describing semi-classical fluctuations
about the minimal surface in AdS and flat spaces, we show that 
it does not depend upon the AdS regularizing parameter $\epsilon$. 
The only dependence upon $\epsilon$ resides in the reparametrization path
integral of the exponential of the classical boundary action.
We analyze how the result depends on the choice of the boundary condition 
imposed on fluctuating fields and show that, despite the 
fact that the contribution of individual angular modes changes,
the product over the modes remains unchanged.
}


\begin{document}

\maketitle 
\setcounter{page}{2}

\tableofcontents

\section{Introduction}

The AdS/CFT correspondence states that the Wilson loop in ${\cal N}=4$ 
$SU(N)$ super Yang--Mills is equal to the disk amplitude of
an open IIB superstring with the Ramond-Ramond flux in the $AdS_5\times S^5$
 background~\cite{Mal98,RY}.
For a circular Wilson loop the supergravity approximation~\cite{BCFM98,DGO99}
to the disk amplitude indeed coincides with 
the limit of large 't~Hooft coupling $\lambda$ of the explicit result for 
the Wilson loop~\cite{ESZ00}. Also the $\lambda$-dependence of 
the pre-exponentials apparently coincides~\cite{DG00}, 
while the comparison of the constant factors relies on the
one-loop computations~\cite{DGT00,SY08,KT08} of the effective 
action for superstring in $AdS_5\times S^5$ 
that involves a nontrivial renormalization
by subtracting the contribution from a reference contour (the straight line).
These constant factors agree only up to a factor of 2, 
which is one of the motivations
to repeat the computation by another method as is done in this Paper.

Circular Wilson loops also emerge in the study of
the Schwinger process of pair production in a constant electric field,
which is calculable in ${\cal N}=4$ super Yang--Mills 
at large $\lambda$ via the AdS/CFT correspondence~\cite{GSS02}. It has 
recently been argued~\cite{SZ11} that at large coupling $\lambda$ there 
exists a critical value of the electric field like in
string theory, contrary to what is the case at
weak coupling. Using a representation
of the string disk amplitude in AdS space through a path integral
over reparametrizations of the boundary, it has recently been shown~\cite{AM11}
that quantum fluctuations about the minimal surface  result in
\be
W({\rm circle})\propto 
\e^{-2\pi \sqrt{\lambda}R /\eps+\sqrt{\lambda}} \lambda^{-3/4}
\left(\frac R\eps\right)^{\nu/2}, 
\label{W1l}
\ee
at the one-loop order. Here $R$ is the radius of the circle
and  $\eps$ is a regularization parameter associated with
moving the boundary of $AdS$ from $Z=0$ to $Z=\eps$, where $Z$ denotes
the radial AdS coordinate. 

The exponent in \eq{W1l} is the classical action, i.e.\ the area of the minimal
surface enclosed by a circle in the boundary \cite{BCFM98,DGO99}.
The induced metric is singular at the boundary and 
the regularization parameter $\eps$
plays in the dual language of  D-branes 
the role of the $U(1)$ boson mass~\cite{Mal98,RY}
\be
m=\frac{\sqrt{\lambda}}{2\pi \del},
\label{Wmass}
\ee 
associated with the breaking 
$U(N)\to U(1)\times U(N-1)$. While it was shown~\cite{DGO99,DF05}
that the $\epsilon$-dependence of the classical action can be eliminated
for (the dual of) 
the Wilson loop by a Legendre transformation, it plays a crucial
role in the computations of the Schwinger effect of production
of a pair of $U(1)$-bosons with masses $m$ given by \eq{Wmass}. 
This is the reason why we shall concentrate in this Paper on the 
dependence of $W({\rm circle})$ upon $\epsilon$.

The pre-exponential factor 
results from quantum fluctuations about the minimal surface.
The factor $\lambda^{-3/4} $ was linked~\cite{DG00} to the 
presence of three $SL(2,\Bbb{R})$ zero modes of the fluctuations
about the classical solution. 
The $\epsilon$-dependence of the pre-exponential
factor displayed in \eq{W1l} 
was obtained in Ref.~\cite{AM11}, where it was argued   
that the value of $\nu$ is expected 
to be 3, conjectured again to be related with the number of
the $SL(2,\Bbb{R})$ zero modes.
Our goal in this Paper will be to reproduce this result, pursuing 
direct computations~\cite{DGT00,SY08,KT08} 
of the one-loop effective action, resulting
from semi-classical quantum fluctuations of 
an open Green--Schwarz superstring in $AdS_5\times S^5$ with the ends at
the boundary circle.

In order to describe our results let us first recall what
the analogous one-loop effective action looks like for
an open bosonic string in flat space. 
In the Polyakov formulation~\cite{Pol81} 
the Liouville field $\varphi$, which emerges 
through the conformal (Weyl) factor in the metric tensor,
\be
g_{ab}=\e^{\varphi} \delta_{ab},
\ee 
decouples in the bulk for the critical dimension $d=26$,
since the conformal anomaly is proportional to $d-26$.
However, its boundary value does not decouple for
off-shell disk amplitudes even in $d=26$ and results in 
a reparametrization path integral
\be
Z_{\rm flat} = \int \D t(s)\, \e^{-K S_{\rm cl}[t(s)]},
\label{Zflat}
\ee
where $K$ is the string tension and 
$S_{\rm cl}[t(s)]$ is a classical boundary action, which emerges
after path-integrating over fluctuations of the open string with
fixed ends, and whose explicit form depends on the choice of the
coordinates parametrizing the string world sheet including its
boundary. The path integral in \eq{Zflat} goes over the functions
$t(s)$, reparametrizing the boundary, with non-negative derivative
$\d t(s)/\d s\geq 0$, which is related to the boundary value of 
the Liouville field $\varphi_{\rm B}$ as
\be
\frac{\d t(s)}{\d s} = \e^{\varphi_{\rm B}/2}.
\ee
While the necessity for 
reparametrizations of the boundary was
emphasized long ago~\cite{Alv83}, only recently some progress
has been achieved~\cite{MO09,BM09} as to how to define the measure
and actually compute the path integral in \eq{Zflat}.

The disk amplitude for the Green--Schwarz string in $AdS_5\times S^5$ 
with the circular boundary can be represented 
at one loop in a similar form
\be
Z_{\rm AdS} = \int \D t(s)\, \e^{-\sqrt{\lambda} S_{\rm cl}[t(s)]}\,
Z^{(1)}_{\rm AdS},
\label{ZAdS}
\ee
where $ S_{\rm cl}[t(s)]$ is explicitly constructed in Ref.~\cite{AM11}
and $Z^{(1)}_{\rm AdS}$ is the ratio of the determinants of second-order operators
explicitly found in Ref.~\cite{DGT00}. By computing this ratio of determinants
we show in this Paper that $Z^{(1)}_{\rm AdS}$ does not depend on the AdS
regularizing parameter $\eps$, so the $\eps$-dependence of the 
one-loop effective action is entirely due to the reparametrization path
integral in \eq{ZAdS}, reproducing the result of Ref.~\cite{AM11}
displayed in \eq{W1l}.

This Paper is organized as follows. In Sect.~\ref{s:pre} we briefly
review the classical solution in AdS for the circular boundary.
In Sect.~\ref{s:res} we describe the main results for the ratio
of determinants, obtained in this Paper. In Sect.~\ref{s:4}
we compute various ratios of 1D determinants and study their
dependence on the choice of the boundary conditions.
In Sect.~\ref{s:eps} we concentrate on the $\eps$-dependence of
the ratios of relevant determinants. In Sect.~\ref{s:sum} we 
compute the ratios of 2D determinants, multiplying 
the 1D determinants over angular modes,
and derive the results listed in Sect.~\ref{s:res}. We also compute
the difference between the contributions from a circle and
a straight line and show that it does not depend on $\eps$.
Sect.~\ref{s:conclu} briefly summarizes the results of this Paper.
Appendices~\ref{appCC} to \ref{appL} are devoted to technical details of
the calculations.

\section{Preliminaries\label{s:pre}}

The upper half-plane parametrization $z=x+\i y$ ($y>0$) 
used in Ref.~\cite{AM11} for constructing the boundary action in AdS space
is not convenient for computing the ratio of determinants for 
circular geometry because the minimal value
$\ym$, associated with the boundary, depends on $x$ for the consistency
with the boundary metric,  as is pointed out there.
For this reason the variables are not separated in a simple way.
It is more convenient to conformally map the upper half-plane 
onto a unit disk and then
the unit disk onto a strip $\x \in [0,\infty)$, $\phi\in [0,2\pi)$ as%
\footnote{A subtlety is that to obtain a unit disk we  
conformally map the upper half-plane with infinity excluded.
Then it has the Euler character one. If alternatively the
upper half-plane is periodically identified along the real axis,
it has topology of a cylinder and the Euler character zero.\label{f:1}}
\be
\om=\e^{-\x+\i \phi}=\frac{\i-z}{\i+z}.
\label{cmap}
\ee
The boundary at $Z=Z_{\rm min}$ now corresponds to
\be
\x_{\rm min}=\del\equiv\frac 12 \ln \frac{R+Z_{\rm min}}{R-Z_{\rm min}}.
\label{defdel}
\ee

The spherical solution of Refs.~\cite{BCFM98,DGO99} for the embedding
space coordinates $Y_{-1}$, $Y_0$, $Y_1$, $Y_2$, $Y_3$, $Y_4$, obeying
\be
Y\cdot Y \equiv -Y_{-1}^2- Y_0^2+Y_1^2+Y_2^2+Y_3^2+Y_4^2=-1,
\label{=1}
\ee 
reads
\begin{subequations}
\bea
Y_1+\i Y_2  &= &\frac{2\om}{1-\om \bar \om}, \\
Y_{-1} &= &\frac{1+\om\bar \om}{1-\om \bar \om},\\
Y_4 &=& Y_0=Y_3=0,
\eea
\label{solu}
\end{subequations}
or
\begin{subequations}
\bea
Z&\equiv&\frac{R}{Y_{-1}-Y_4}=R\frac{1-\om \bar \om}{1+\om \bar \om} ,\\
X^1 +\i X^2&\equiv& Z (Y_1+\i Y_2)=R\frac{2\om}{1+\om \bar \om},
\eea
\label{PP}
\end{subequations}
on the Poincare patch, so the induced metric
\be
\d \ell^2 = \frac{\d \om \d \bar \om}{(1- \om\bar \om)^2},
\label{Poin}
\ee 
is the metric of the Lobachevsky plane for the Poincare disk.
The solution~\rf{PP} describes a sphere 
$(X^1)^2+(X^2)^2+Z^2=R^2$ and
corresponds to a circle of the radius $R$ in the boundary when $Z=0$.  

For the coordinates~\rf{cmap} the solution~\rf{PP} and the metric~\rf{Poin} 
read
\be
Z=R \tanh \x,\qquad X^1+\i X^2 =\frac{R \e^{\i \phi}}{\cosh\x},
\qquad X^0=X^3=0,
\label{0sol}
\ee 
and
\be
\d \ell^2 = \frac{1}{\sinh ^2\x}\left( \d \x^2+\d \phi^2  \right).
\label{Poiny}
\ee

The solution \rf{solu} obeys the Euler--Lagrange equation 
\be
(-\Delta+2)Y_i=0,\qquad
\Delta=(1-\om\omb)^2 \frac{\partial^2}{\partial \om \partial \omb},
\ee
or
\be
\Delta= \sinh^2\x \left( \frac{\partial^2}{\partial \x^2 }
+ \frac{\partial^2}{\partial \phi^2} \right),
\ee
for the coordinates~\rf{cmap}.
On the Poincare patch it takes the form 
\begin{subequations}
\bea
\partial_a \frac1{Z^2} \partial_a X^\mu=0, 
\label{ec1}\\*
\partial_a \frac1{Z^2} \partial_a Z+2\frac{\sqrt{g}}{Z}=0, 
\label{ec2}
\eea
\end{subequations}
where 
\be
\sqrt{g}=\frac{1}{\sinh^2\x}.
\ee

As usual, the $SL(2,\Bbb{R})$ transformation is an isometry of this
spherical solution.
The $SL(2,\Bbb{R})$ coordinate transformation of the upper
half-plane reads
\be
z\to \frac{a z+b}{c z +d},\quad a d-bc =1,
\label{Moble}
\ee
with real $a$, $b$, $c$ and $d$ to preserve the boundary (= real axis).
After the conformal mapping \rf{cmap} it takes the form
\be
\om\to \e^{\i \alpha}\frac{\om- \om_0}{1- \om \bar \om_0},
\label{UDmap}
\ee
with real $\alpha$ and complex $\om_0$
($\bar \om_0$ denotes complex conjugation), which maps a unit disk onto itself.


Finally, we mention that for small $\x$ the solution \rf{0sol} for a circle
reproduces near the boundary the one for a straight line
\be
X^1=x, \qquad X^2=X^3=X^4=0, \qquad Z=y .
\label{stra}
\ee
For the former case the strip coordinates are more convenient, while for the 
latter case the upper half-plane coordinates are more convenient.

\section{Results for the ratio of determinants\label{s:res}}

The ratio of one-loop determinants that enter \eq{ZAdS} 
reads explicitly~\cite{DGT00}
\be
Z^{(1)}_{\rm AdS}=
\frac{\det\left(-\Delta_{ij}+\delta_{ij}\right)^{1/2}_{\rm ghost}}
{\det\left(-\Delta_{ij}+\delta_{ij}\right)^{1/2}_{\rm long.}}
\frac{\det\left(-\widehat\nabla^2+R^{(2)}/4+1 \right)^{8/2}_{\rm Fermi}}
{\det\left(-\Delta+2 \right)^{3/2}_{\rm Bose}
\det\left(-\Delta \right)^{5/2}_{\rm Bose}}.
\label{therati}
\ee
Here $\Delta =\nabla^2=
 g^{ab}\nabla_a \nabla_b $ stands for the Laplacian in general coordinates
for an appropriate representation,
and the masses (squared) equal 2 for transverse bosons, 
1 for longitudinal bosons and 1 for fermions. 
They emerge because of the curvature of the embedding (AdS) space.
The operators in the ghost and longitudinal determinants 
are the same, but the ratio is generically not 1 because
of different boundary conditions.

Our strategy to evaluate the $\eps$-dependence of the ratio~\rf{therati}
is to utilize the fact that the analogous ratio of the determinants in
flat space does obviously not depend on $\eps$,
so we assume that $Z^{(1)}_{\rm flat}=1$ as is expected from
supersymmetry and calculate the ratio
\be
\frac{Z^{(1)}_{\rm AdS}}{Z^{(1)}_{\rm flat}}=
\frac{\det\left(-\Delta\right)}
{\det\left(-\Delta_{ij}+\delta_{ij}\right)^{1/2}}
\left[\frac{\det\left(-\widehat\nabla^2+R^{(2)}/4+1 \right)}
{\det\left(-\widehat\nabla^2+R^{(2)}/4 \right)}\right]^{8/2}
\left[\frac{\det\left(-\Delta \right)}{\det\left(-\Delta+2 \right)}
\right]^{3/2},
\label{thefla}
\ee
of massive to massless (associated with flat space) determinants,
noting that the ghost determinants are the same.
This is in contrast to the proposal~\cite{DGT00} to evaluate 
the ratio of the $Z^{(1)}_{\rm AdS}$'s for the circle and
the straight line, where the $\eps$-dependence, we are interested in, 
cancels out.

The massive determinants are not directly computable  by
the Seeley coefficients, widely used in the 1980's for  the massless case,
 because the variation with respect to the metric
is not reduced to an anomaly. 
We employ instead the method advocated
for this problem by Kruczenski and Tirziu~\cite{KT08}, which is
based on direct computation of 1D$\times$angular determinants,
applying the Gel'fand--Yaglom technique for the 1D determinants. 
This is possible because all determinants in \eq{thefla} are to be
calculated for the Dirichlet boundary condition.

The ratio~\rf{thefla} crucially simplifies if we compute it for
a straight line rather than for a circle. This is legitimate since
we are interested only in the $\epsilon$-dependence of the ratio~\rf{thefla},
which originates from the region of $\x\sim\eps$ near the boundary,
where the solution~\rf{0sol} for a circle can be substituted by 
the solution~\rf{stra} for a straight line.
For the bosonic and fermionic determinants this was explicitly
demonstrated by computations in Ref.~\cite{KT08}. Of course this
is not the case for $\eps$-independent constants in the determinants, 
which are different for the circle and the straight line. 

One may wonder if one can lose the $SL(2,\Bbb{R})$ zero modes when replacing
the circle by the straight line? The answer is ``no'', because 
the $SL(2,\Bbb{R})$ zero modes show up in the determinant of
the ghost operator which is
the same in AdS and flat space and therefore cancels in the ratio~\rf{thefla}. 

The Gel'fand--Yaglom technique expresses the ratio of 1D determinants
through the (properly normalized) solutions of the equations
\be
\left(-\partial^2+V_i(\x) \right) f_i(\x)=0, 
\qquad f_i(\eps)=0,~~ 
f^\prime_i(\eps)=1 \quad i=1,2,
\ee
as
\be
\frac{\det \left(-\partial^2+V_1(\x)\right)}
{\det \left(-\partial^2+V_2(\x)\right)}=
\frac{f_1(\infty)}{f_2(\infty)}.
\ee

Applying this technique, we obtain the following results. 
The ratio of the massive to massless bosonic determinants is explicitly 
\be
\prod_\om\frac{\det^{3/2}\, (-\partial^2+\om^2+2/\x^2)}
{\det^{3/2}\,(-\partial^2+\om^2)}
=\prod_\om\left( 1+\frac1{\eps\om} \right)^3=
\e^{\frac 3\eps \left[\ln(\Lambda\eps)+1\right]}. 
\ee
The ratio of the massive to massless fermionic determinants is explicitly%
\footnote{Here ${\rm Ei}(-x)$ is the exponential integral.}
\bea
\lefteqn{\hspace*{-1cm} 
\prod_\om \frac{\det^{4/2}\, (-\partial^2+\om^2+3/4\x^2+\om/\x)
\det^{4/2}\, (-\partial^2+\om^2+3/4\x^2-\om/\x)}
{\det^{4/2}\, (-\partial^2+\om^2-1/4\x^2+\om/\x)
\det^{4/2}\, (-\partial^2+\om^2-1/4\x^2-\om/\x)} } \non
&=&\prod_\om\left( \frac{1+\frac1{2\eps\om}}
{-2\eps\om \e^{-2\eps\om} {\rm Ei}(-2\eps\om)} \right)^4=
\e^{\frac 4\eps \left[\ln(\Lambda\eps)+\frac12+C_1 \right]}.
\eea
The ratio of the massive to massless longitudinal determinants is explicitly 
\bea
\lefteqn{\hspace*{-1cm} 
\prod_\om\frac{\det^{1/2}\, (-\partial^2+\om^2+2/\x^2+2\om/\x)
\det^{1/2}\, (-\partial^2+\om^2+2/\x^2-2\om/\x)}
{\det \,(-\partial^2+\om^2)}} \non
&=&\prod_\om\left( 1+\frac1{\eps\om}+\frac1{2\eps^2\om^2} \right)=
\e^{\frac 1\eps\left[ \ln(\Lambda\eps)+1+\pi/4+1/2 \ln2 \right]}.
\eea
Multiplying these three, we finally find
\be
\frac{Z^{(1)}_{\rm AdS}}{Z^{(1)}_{\rm flat}}=
\e^{ (4-3-1)\frac 1\eps \ln(\Lambda\eps)+{C_2}/\eps}=
\e^{{C_2}/\eps}.
\label{C2}
\ee

Our results differ from those of Ref.~\cite{KT08},
where the ratio of the ghost to longitudinal determinants was assumed to be 1. 
For our results the $\frac1\eps\ln\eps$ term coming from the bosonic 
and fermionic determinants is precisely canceled by the one
coming from the ratio of the longitudinal determinants. 
The cancellation is as for the $\frac1\eps \ln\Lambda$ divergent parts:
\be
2\times 1 \;\hbox{(longitudinal)} + 3\times 2\; \hbox{(transversal)} - 
8\times1 \;\hbox{(GS fermions)}=0,
\ee 
where the first figure in each term is the number of degrees
of freedom and the second one is the proper mass squared.
The remaining
term is $1/\eps$, which does not spoil anything and is removable
by the Legendre transformation like the classical singularity.
It can be simply viewed as 
a renormalization of the $U(1)$ boson mass~\rf{Wmass}.

Therefore $Z^{(1)}_{\rm AdS}$ is equal to a constant
that does not depend on $\eps$ after the Legendre transformation 
and is not essential in \eq{ZAdS}.
Like in the flat space the Liouville field $\varphi(x,y)$ 
($g_{ab}=\e^{\varphi} \delta_{ab}$) decouples in the bulk, while its boundary
value is related to the reparametrizing function $t(s)$ as
\be
\frac{\d t(s)}{\d s}=\e^{\varphi(s,\eps)/2}.
\ee
We are thus left with the same boundary action $ S_{\rm cl}[t(s)]$ 
as obtained in Ref.~\cite{AM11} 
(an extension of Douglas' integral~\cite{Dou31} to AdS space), 
which is to be substituted in the reparametrization path integral, 
reproducing the effective action displayed in \eq{W1l}.

In the rest of this Paper we present technicalities used for
the derivation of this result.

\section{Gel'fand-Yaglom meets Gel'fand-Dikii\label{s:4}}

As is already pointed out, the Gel'fand--Yaglom technique is 
applicable when the ratio of determinants is calculated for
the Dirichlet boundary condition. In Ref.~\cite{KT08} it was
imposed at $\x=\eps$, as inherited from the regularization of
the classical action. However, there is nothing special about this
point for the classical solution \rf{solu} or \rf{0sol}, which obeys
the boundary condition at $\x=0$ rather than at $\x=\eps$.
One can therefore wonder if the results will change when we 
impose the Dirichlet boundary condition at $\x=0$ rather
than $\x=\eps$ and consider $\eps$ only  as a parameter regularizing
determinants for the given (Poincare) metric.

Using a more general technique, to be introduced shortly, we find that 
the answer to this question is that
the 1D determinants will change only in a way that results in the same
$\eps$-dependence.
In addition, we confirm the results obtained by
the Gel'fand--Yaglom technique by this other method.

\subsection{Determinants via diagonal resolvent \label{s:dia}}

The ratio of the determinants of two Schr\"odinger operators
of the form $-\partial^2+\om^2+V(\x)$ can be related to the diagonal
resolvent
\be
R_\om(\x,\x;V)
\equiv \LA \x \left|\frac{1}{-\partial^2+\om^2+V(\x)}\right|\x\RA, 
\label{di}
\ee
where $\om^2$ is a spectral parameter and $\partial\equiv\partial/\partial \x$,
as follows
\be 
{\cal R}_\om\equiv
\frac{{\rm det}\left(-\partial^2 + \om^2 + V(\x)  \right)}
{{\rm det}\left(-\partial^2 + \om^2   \right)} 
= \exp\left[\int \d \om^2 \int_\del^\infty \d \x\, 
\left(R_\om(\x,\x;V)-R_\om(\x,\x;0) \right)\right]. 
\label{GD}
\ee
Here an overall constant is to be fixed by requiring that the ratio tends to
1 as $\om\to\infty$, since 
the potential can then be disregarded.
We have introduced $\eps$ as a lower limit 
of the integral over $\x$,
anticipating a divergence at small $\x$.

The diagonal resolvent $R_\om(\x,\x;V)$ can be easily constructed via
two solutions of the second-order equation
\be
\left(-\partial^2+\om^2+V(\x) \right)f_\pm(\x)=0,
\label{s-o}
\ee
where $f_+(\x)$ vanishes as $\x\to\infty$ and $f_-(\x)$ obeys
the boundary condition at the beginning of the interval. This can
be either the Dirichlet or Neumann or mixed (Robin) boundary condition.
The third case is needed for the ghost determinant~\cite{Alv83,DOP82,Luc89}. 
The explicit formula is well known:
\be
R_\om(\x,\x;V)
=\frac{f_+(\x) f_-(\x)}{f_+(\x)\partial f_-(\x) -f_-(\x)\partial f_+(\x)}.
\label{dia}
\ee
It is less known that this resolvent obeys the quadratic 
Gel'fand--Dikii equation
\be
-2R_\om \partial^2 R_\om +(\partial R_\om)^2+4 (\om^2+V)R_\om^2=1,
\label{GDe}
\ee 
which may help to find it even when it is difficult to solve 
\eq{s-o} explicitly.
This method of computing the ratio of determinants 
is described in more detail
in Ref.~\cite{Mak02} for the case of fluctuations about an instanton 
for the double-well potential. 
 
Finally, to compute the (logarithm of the) ratio of 2D determinants we have to 
sum over angular modes for a circular boundary:
\be 
\ln \frac{{\rm det}\left(-\partial^2_a  + V(\x)  \right)}
{{\rm det}\left(-\partial^2_a    \right)}
=\sum_\om \int \d \om^2 \int_\del^\infty \d \x\, 
\left(R_\om(\x,\x;V)-R_\om(\x,\x;0) \right),
\label{2Ddet}
\ee
where $\om$ runs over integers or half-integers for
bosonic or fermionic determinants, respectively.
For a straight line the sum over $\om$ in \eq{2Ddet}
is to be replaced by an integral.

\subsection{Bosonic determinant}

For the bosonic determinant the potential is~\cite{DGT00}
\be
V_b(\x)= \frac{2}{\sinh^2 \x}.
\label{Vb}
\ee
To make a connection with the results of Ref.~\cite{KT08}, we begin with
the case, where $f_-(\x)$ vanishes at $\x=\x_0$. We shall then set $\x_0$
either to $\eps$, reproducing the results of Ref.~\cite{KT08},
or to 0, answering the question posed at the beginning of this Section.

The two solutions to \eq{s-o} with the potential \rf{Vb} are 
\begin{subequations}
\bea
f_+(\x)&= &(\coth \x +\om )\e^{-\om \x}, \label{42a}\\
f_-(\x)&=&(\coth \x -\om )\e^{\om \x} - 
(\coth \x +\om) 
\e^{\om(2\x_0- \x)}
\frac{(\coth \x_0 -\om)}{
(\coth \x_0 +\om)},  \label{42b} 
\eea
\label{bsolu1}
\end{subequations}
where $f_-(\x_0)=0$ so that $R_\om(\x,\x;V_b)$ obeys the boundary condition
$R_\om(\x_0,\x_0;V_b)=0$. Then we have
\be
R_\om(\x,\x;V_b)=
\frac{(\om+\coth\x)}{2\om \left(\om^2-1\right)}
\Big[(\om-\coth\x) 
-\e^{2\om (\x_0-\x)} (\om+\coth\x)\frac{(\om-\coth\x_0)}
{(\om+\coth{\x_0})}\Big],
\ee
which satisfies \eq{GDe}.

Analogously in the free case of $V=0$ we find
\begin{subequations}
\bea
f_+(\x)&=&\e^{\om(\x_0-\x)},\\
f_-(\x)&=&\frac{\sinh{\om(\x_0-\x)}}{\om},
\label{fs1b}
\eea
\label{freesolu1}
\end{subequations}
so the free resolvent is
\be
R_\om(\x,\x;0)=\frac{1-\e^{2\om (\x_0-\x)}}{2 \om},
\label{freeres}
\ee
reproducing the usual one when $\x_0\to-\infty$.
Equation~\rf{freeres} will be extensively used below,
when evaluating the ratios of determinants

These $f_-(\x)$'s shown in Eqs.~\rf{42b} and \rf{fs1b} vanish at $\x=\x_0$, but
the lower limit in the integral~\rf{GD} which we denote again as 
$\eps$ could be greater then $\x_0$ ($\eps \geq \x_0$). 
In this case the solutions do not have zeros for $\x>\del$, so
$\del$ plays simply the role of a regularization which is not
directly related to the boundary condition. We can thus
check to what extent the obtained results will be independent of 
$\x_0$, i.e. of the choice of  the boundary condition. 

For the ratio of the determinants \rf{GD} we obtain
\bea
{\cal R}_\om 
&=&
\exp\left[\frac{1}{2} (\coth\eps-1) \Big( \text{Ei}[-2(\om-1)(\eps-\x_0)]
-\ln(\om-1) + \ln(\om+1) \Big) \right. \non 
&&\hspace*{1cm} -
\frac 12(\coth\eps+1) \text{Ei}[-2(\om+1)(\eps-\x_0)]\non
&&\hspace*{1cm} \left.
-\e^{2 (\eps-\x_0) \coth\x_0} \text{Ei}[-2(\om+\coth\x_0)(\eps-\x_0)]
\frac{\sinh(\eps-2\x_0)}{\sinh\eps }\right],
\label{77}
\eea
where the exponential integral
\be
 \text{Ei}(-x)\equiv-\int_x^\infty \frac{\d t}t \e^{-t} 
\stackrel{x\to0} \longrightarrow \ln x +\gamma_{\rm E}-x +{\cal O}(x^2).
\label{ei}
\ee

A few comments concerning \eq{77} are in order. When $\x_0\to \eps$,
it gives by the use of the asymptote~\rf{ei}
\be 
\left.{\cal R}_\om \right|_{\x_0=\eps}=
\frac{\om + \coth \del}{\om+1},
\label{good}
\ee
reproducing the result of Ref.~\cite{KT08} obtained by 
the Gel'fand--Yaglom technique.
For $\om\to+\infty$ the exponent in \eq{77} vanishes,
so the ratio of determinants tends to 1 as it should. 
Another interesting case 
is when $\x_0\sim\eps$ but still $\x_0<\eps$, say  $\x_0=\eps/2$.
Then the term displayed in the third line of \eq{77} can be disregarded.
The same is true for $\x_0=0$, when
\bea
\left.{\cal R}_\om \right|_{\x_0=0}&=&
\exp\left[\frac{1}{2} (\coth\eps-1) \left( \text{Ei}[-2(\om-1)\eps)]
-\ln(\om-1)+\ln(\om+1) \right)\right. \non 
&&\hspace*{1cm} \left.-
\frac 12(\coth\eps+1) \text{Ei}[-2(\om+1)\eps] \right].
\label{79}
\eea
As $\om\to+\infty$ this tends to 1, but behaves at large $\omega$ smoother 
than \rf{good}.

If alternatively $\om \ll 1/\eps$ as $\eps\to0$, we get
\be
\ln{\cal R}_\om \stackrel{\eps\to0}\longrightarrow -\left[\ln(2\eps)
+\gamma_{\rm E}-2+\ln(\om+1)\right].
\label{48}
\ee
This has the same $\eps$-dependence as $\eps\to0$ limit of
the (logarithm of the)
right-hand side of \eq{good}, but differs by a constant.

\subsection{Fermionic determinant}

The fermionic potential reads~\cite{DGT00,SY08}
\be
V_{f\pm}(\x)= \frac{3}{4\sinh^2 \x} + \frac 14 \pm \om  \coth \x,
\label{Vf}
\ee
where the $+(-)$ sign refers to positive (negative) frequencies $\om$.

The treatment of the fermionic determinant 
follows that of the bosonic ones except for two major differences:

1) Anti-periodicity requires the angular modes to be $\e^{\i r \phi}$
with half-integer $r \in \Bbb{Z}+1/2$.

2) The fermionic potential $V_f$ defined in \rf{Vf} depends on $\omega$ itself.
For this reason we have
\bea 
{\cal R}_{f\pm}&\equiv&
\frac{{\rm det}\left(-\partial^2 + \om^2 + V_{f\pm}(\x)  \right)}
{{\rm det}\left(-\partial^2 + (\om\pm \frac12)^2   \right)} \non 
&= &\exp\left\{\int \d \om \int_\del^\infty \d \x\, 
\left[(2\om\pm \coth\x)R_\om(\x,\x; V_{f\pm})
-(2\om\pm1) R_{\om\pm1/2}(\x,\x;0) \right]\right\}, \non &&
\label{GDf}
\eea
that generalizes \eq{GD} to the case of such an $\om$-dependent potential.
Also, the cases of positive and negative $\omega$ should now be treated
separately, so the $\pm$ in \eq{GDf} refer to positive or negative 
frequencies $r$.

For positive half-integer $\om=r>0$ the two solutions to \eq{s-o} 
with the potential~\rf{Vf} are
\begin{subequations}
\bea
f_+(\x)&=& \frac{\e^{-\om \x}}{\sqrt{\sinh \x}}, \\*
f_-(\x)&=& \frac{\e^{\om \x}\left(-\cosh \x+2 \om \sinh \x \right)}
{\sqrt{\sinh \x}} 
-\frac{\e^{\om(2\x_0- \x)}\left(-\cosh \x_0+2 \om \sinh \x_0 \right)}
{\sqrt{\sinh \x}} .~~~
\eea
\end{subequations}
Calculating the diagonal resolvent~\rf{dia}, substituting into \eq{GDf}
and integrating, we obtain for the ratio 
\bea
\ln {\cal R}_{f+}&=& 
\frac14 \left( \coth\eps-1 \right)
\Big\{ {\rm Ei} \left[-(2\om-1)(\eps-\x_0)\right] -
{\rm Ei} \left[-(2\om+1)(\eps-\x_0)\right]
\non && \hspace*{3cm}-\ln(2\om-1)+\ln(2\om+1)
\Big\} + \, C,
\label{77f+}
\eea
where $C$ is an integration constant that does not depend on $\om$.
It is given by
\be
C=\frac 12 \ln \frac{\coth\eps+1}2,
\label{C}
\ee
as is derived in Appendix~\ref{appCC} using the semi-classical expansion.

For negative half-integer $r=-\om<0$  the two solutions to \eq{s-o} 
with the potential~\rf{Vf} are
\begin{subequations}
\bea
f_+(\x)&=& \frac{\e^{-\om \x}\left(\cosh \x+2 \om \sinh \x \right)}
{\sqrt{\sinh \x}}, \\*
f_-(\x)&=& \frac{\e^{\om \x}}{\sqrt{\sinh \x}} 
-\frac{\e^{\om(2\x_0- \x)}\left(\cosh \x+2 \om \sinh \x \right)}
{\sqrt{\sinh \x}\left(\cosh \x_0+2 \om \sinh \x_0 \right)} .
\eea
\end{subequations}
The ratio \rf{GDf} of the determinants reads
\bea
\ln {\cal R}_{f-} &=&
\frac{1}{4} (\coth\eps-1) \Big\{ \text{Ei}[-(2\om-1)(\eps-\x_0)]
-\ln(2\om-1)+\ln(2\om+1) \Big\} \non 
&& 
- \frac 14\left(\coth\eps+3 \right) \text{Ei}[-(2\om+1)(\eps-\x_0)]\non
&& 
-\frac14 \e^{(\eps-\x_0) \coth\x_0} 
\text{Ei}[-(2\om+\coth\x_0)(\eps-\x_0)]
\frac{1+\cosh[2(\eps-\x_0)]-2\cosh(2\x_0)}{\sinh\eps \sinh\x_0} \non
&& -\,C,
\label{77f-}
\eea
with the same $C$ as in \eq{77f+} given by \eq{C}.

In the limit $\x_0\to \eps$ we have
\be
\left.\ln {\cal R}_+ \right|_{\x_0=\eps}=C,
\ee
and
\be
\left.\ln {\cal R}_-\right|_{\x_0=\eps}= \ln (2\om+\coth\eps)-\ln (2\om+1)-C,
\ee
reproducing the results of Ref.~\cite{KT08}. Alternatively, for $\x_0=0$,
we obtain
\bea
\left.\ln {\cal R}_{f+}\right|_{\x_0=0}&= &
\frac14 \left( \coth\eps-1 \right)
\Big\{ {\rm Ei} \left[-(2\om-1)\eps\right] -
{\rm Ei} \left[-(2\om+1)\eps\right]\non 
&& \hspace*{2.7cm}
-\ln(2\om-1)+\ln(2\om+1)
\Big\}+ \, C,
\label{77f+0}
\eea
and
\bea
\left.\ln {\cal R}_{f-}\right|_{\x_0=0} &=&
\frac{1}{4} (\coth\eps-1) \Big\{ \text{Ei}[-(2\om-1)\eps]
-\ln(2\om-1)+\ln(2\om+1) \Big\} \non 
&& 
- \frac 14\left(\coth\eps+3 \right) \text{Ei}[-(2\om+1)\eps]
+\frac {\sinh\eps}{2\eps} \e^{-2\om \eps} -\,C.
\label{77f-0}
\eea

For the product of these two ratios (one with positive and one with negative
$\om$) we obtain [cf.\ \eq{77} for bosons]
\bea
\lefteqn{  {\cal R}_{f+} {\cal R}_{f-} =
\exp\left[\frac{1}{2} (\coth\eps-1) \Big\{ \text{Ei}[-(2\om-1)(\eps-\x_0)]
-\ln(2\om-1)+\ln(2\om+1) \Big\}\right. 
}
\non 
&& 
- \frac 12\left(\coth\eps+1\right) \text{Ei}[-(2\om+1)(\eps-\x_0)]\non
&& 
\left. -\frac14 \e^{(\eps-\x_0) \coth\x_0} 
\text{Ei}[-(2\om+\coth\x_0)(\eps-\x_0)]
\frac{1+\cosh[2(\eps-\x_0)]-2\cosh(2\x_0)}{\sinh\eps \sinh\x_0}\right].
\hspace*{-1cm}\non
&&
\label{77f}
\eea

If $\x_0\to\eps$, we can use the expansion \rf{ei} to find 
\be
\left.{\cal R}_{f+} {\cal R}_{f-}\right|_{\x_0=\eps}=
\frac{2\om +\coth\eps}{2\om+1},
\label{67}
\ee
reproducing the result of Ref.~\cite{KT08} obtained by the other method.
Alternatively, for $\x_0=0$ the term displayed in the third line of
\eq{77f} simplifies and we get
\bea
\left.{\cal R}_{f+} {\cal R}_{f-}\right|_ {\x_0=0}&=&
\exp\left[\frac{1}{2} (\coth\eps-1) \Big\{ \text{Ei}[-(2\om-1)\eps]
-\ln(2\om-1)+\ln(2\om+1) \Big\}\right. \non 
&& \hspace*{8mm} 
\left.- \frac 12\left(\coth\eps+1\right) \text{Ei}[-(2\om+1)\eps]+ 
\frac{\sinh\eps}{2\eps} \e^{-2\om\eps} \right].
\label{79f}
\eea
This is to be compared with \eq{79} in the bosonic case.

\section{Straight-line limit ($\boldsymbol{\epsilon}$-dependence)\label{s:eps}}

As is already pointed out, the case of a straight line can be obtained 
from the case of a circular boundary when distances from the 
boundary are of the order of $\eps$. This implies that we are dealing 
with the $\om\sim 1/\eps$ limit. The formulas of 
this Section can be obtained as this limit of the proper formulas
for a circle. 

\subsection{Massive bosons and fermions}

For the bosonic determinant with $\om=|m|$
we obtain from \eq{77}
\bea
\ln\frac{\det (-\partial^2+\om^2+2/\x^2)}{\det( -\partial^2+\om^2)}=
\frac{1-\e^{-2\om(\eps-\x_0)}}{\om \eps}-{\rm Ei}[-2\om(\eps-\x_0)] \non  -
\e^{2(\eps-\x_0)/\x_0}\left(1-\frac{2\x_0}{\eps}\right)
{\rm Ei}\left[-2\left(\om+\x_0^{-1}\right)(\eps-\x_0) \right].
\label{sdb}
\eea

For the fermionic determinants with positive $r=\om$ we analogously obtain
from \eq{77f+}
\bea
\ln\frac{\det (-\partial^2+\om^2+3/(4\x^2)+\om/\x)}{\det( -\partial^2+\om^2)}=
\frac{1-\e^{-2\om(\eps-\x_0)}}{4\om \eps}+ C.
\label{sdf+}
\eea

For the fermionic determinant with negative $r=-\om$ we analogously obtain
from \eq{77f-}
\bea
\ln\frac{\det (-\partial^2+\om^2+3/(4\x^2)-\om/\x)}{\det( -\partial^2+\om^2)}=
\frac{1-\e^{-2\om(\eps-\x_0)}}{4\om \eps}
-{\rm Ei}[-2\om(\eps-\x_0)] \non  +
\e^{(\eps-\x_0)/\x_0}\left(-\frac{\eps}{2\x_0}+1+\frac{\x_0}{2\eps}\right)
{\rm Ei}\left[-\left(2\om+\x_0^{-1}\right)(\eps-\x_0) \right]- C.
\label{sdf-}
\eea

When $\x_0=\eps$ we get from Eqs.~\rf{sdb}, \rf{sdf+} and \rf{sdf-}
\begin{subequations}
\bea
\left.\rf{sdb}\right|_{\x_0=\eps}&=& \ln(\om+\eps^{-1})-\ln \om, 
\label{ssb}\\*
\left.\rf{sdf+}+\rf{sdf-}\right|_{\x_0=\eps}&= &\ln(\om+\eps^{-1}/2)-\ln \om,
\label{ssf}
\eea
\end{subequations}
reproducing the results of Ref.~\cite{KT08}.

Alternatively, for $\x_0=0$ we get from Eqs.~\rf{sdb}, \rf{sdf+} 
and \rf{sdf-}
\begin{subequations}
\bea
\left.\rf{sdb}\right|_{\x_0=0}&=& 
\frac{1-\e^{-2\om\eps}}{\om \eps}-{\rm Ei}[-2\om\eps] , 
\label{ssb0}\\*
\left.\rf{sdf+}+\rf{sdf-}\right|_{\x_0=0}&= &
\frac{1-\e^{-2\om\eps}}{2\om \eps}
-{\rm Ei}[-2\om\eps] +\frac12 \e^{-2\om\eps},
\label{ssf0}
\eea
\end{subequations}
reproducing the limits of Eqs.~\rf{79}, \rf{79f}.

\subsection{Longitudinal modes}

The two longitudinal modes have mass 1 and the corresponding 
ratio of determinants reads explicitly~\cite{DGT00}
\bea
\lefteqn{\frac
{\det\left(-\Delta_{ij}+\delta_{ij}\right)^{1/2}}
{\det\left(-\Delta\right)} } \non &&=\prod_{\om}
 \frac{\det^{1/2}\, (-\partial^2+\om^2+2/\x^2+2\om/\x)
\det^{1/2}\, (-\partial^2+\om^2+2/\x^2-2\om/\x)}
{\det \,(-\partial^2+\om^2)}, 
\label{ld}
\eea
which can be treated similarly to the massive fermionic determinants
in the previous Subsection.

The two independent solutions, obeying the Dirichlet boundary condition
at $\x=\x_0$ and $\x=\infty$, are 
\begin{subequations}
\bea
f_+(\x)&=&\frac{\e^{-\om \x}}{2 \om \x},\\
f_-(\x)&=&\frac{\e^{-\om (\x_0+\x)} \left[\e^{2 \om \x} 
(1+2 \om \x (\om \x-1))-\e^{2 \x_0 \om} 
(1+2 \x_0 \om (\x_0 \om-1))\right]}{4 \x_0 \om^3 \x},
\label{fmp}
\eea
\end{subequations}
for $\om>0$ and
\begin{subequations}
\bea
f_+(\x)&=&\frac{\e^{-\om \x} (1+2 \om \x (1+\om \x))}{\om \x}, \\
f_-(\x)&=&\frac{\e^{-\om (\x_0+\x)} 
\left[\e^{2 \om \x} (1+2 \x_0 \om (1+\x_0 \om))-\e^{2 \x_0 \om} 
(1+2 \om \x (1+\om \x))\right]}
{4 \x_0 \om^3 \x},
\label{fmm}
\eea
\end{subequations}
for $\om<0$. 

The solutions~\rf{fmp} and \rf{fmm} are properly normalized to 
be used in the Gel'fand--Yaglom technique when $\x_0=\eps$.
We thus obtain
\be
\rf{ld}= \prod_\om  \left(1+\frac1{\eps\om}+\frac1{2 \eps^2 \om^2}\right).
\label{ldf}
\ee

Analogously, we can perform the computation of the ratio~\rf{ld} by
the Gel'fand--Dikii method using the formula
\be
\frac{{\rm det}\left(-\partial^2 + \om^2 \pm V_{l\pm} \right)}
{{\rm det}\left(-\partial^2 + \om^2   \right)}
= \exp\left\{\int \d \om \int_\del^\infty \d \x\, 
\left[(2\om\pm 2/\x)R_\om(\x,\x; V_{l\pm})
-2\om R_{\om}(\x,\x;0) \right]\right\},
\label{GDl}
\ee
with
\be
 V_{l\pm}= \frac2{\x^2} \pm \frac {2 \om}{\x}, 
\label{Vpm}
\ee
which is an analog of \eq{GDf}, since the potential \rf{Vpm}
is also $\om$-dependent.
For $\x_0=\eps$ this reproduces \eq{ldf}.

For $\x_0=0$ we find from \eq{GDl}
\be
\ln\rf {ld}=\int \d \om \,
\left[\frac1{\eps\om} + \e^{-2 \eps \om} 
\left(\frac54 - \frac{1}{\eps\om} + \frac{1}{2} \eps \om\right) - 
\frac{1}{2} {\rm Ei}(-2 \eps \om)\right].
\label{ldf0}
\ee

\subsection{Massless fermions} 

The last what remains to compute is the ratio of the determinants of
massless fermions to bosons:
\be
\frac{\det\left(-\widehat\nabla^2+R^{(2)}/4 \right)}
{\det\left(-\Delta \right)}=
\prod_\om\frac{\det\, (-\partial^2+\om^2-1/4\x^2+\om/\x)
\det\, (-\partial^2+\om^2-1/4\x^2-\om/\x)} 
{\det^2\, (-\partial^2+\om^2) },
\label{mfd}
\ee
which enters the ratio \rf{thefla}.

The two independent solutions are 
\begin{subequations}
\bea
f_+(\x) &= &\e^{\om \x} \sqrt{x} {\rm Ei}(-2 \om \x),\\
f_-(\x)& =& \e^{\om \x} \sqrt{
  x} ( {\rm Ei}(-2 \om \x) -  {\rm Ei}(-2 \om \x_0)) \sqrt{\x_0}
 \e^{ \om \x_0},
\label{mfp}
\eea
\end{subequations}
for $\om>0$ and
\begin{subequations}
\bea
f_+(\x) &= &\e^{-\om \x} \sqrt{x},\\
f_-(\x) &= &\e^{-\om \x} \sqrt{x} 
( {\rm Ei}(2 \om \x)-  {\rm Ei}(2 \om \x_0)) \sqrt{\x_0}
 \e^{ -\om \x_0},
\label{mfm}
\eea
\end{subequations}
for $\om<0$.

The solutions~\rf{mfp} and \rf{mfm} are properly normalized to 
be used in the Gel'fand--Yaglom technique when $\x_0=\eps$.
We thus obtain
\be
\rf{mfd}=-\prod_\om 2 \e^{2 \eps \om} \eps \om {\rm Ei}(-2 \eps \om).
\label{mfd13}
\ee

Analogously, we can perform the computation of the ratio~\rf{mfd} by
the Gel'fand--Dikii method using the straight line limit of \eq{GDf}.
For $\x_0=\eps$ this reproduces \eq{mfd13}.

For $\x_0=0$ we obtain 
\be
\rf{mfd}= \prod_\om
\frac 12 \left[(2 + \e^{2 \om \eps}) {\rm Ei} (-2 \om \eps) - 
   \e^{-2 \om \eps} {\rm Ei}(2 \om \eps)  \right].
\label{mfd0}
\ee

\section{Multiplying over angular modes\label{s:sum}}

We have described in the two previous Sections how to calculate 1D
determinants. Our primary goal is to use the results
to evaluate the ratio of 2D determinants of the type
\be
{\cal R}=\frac{\det(-\Delta+\mu^2)}{\det(-\Delta)}.
\label{calR}
\ee
We deal with the case, where the Weyl factor of the metric
in conformal coordinates $\x$ and $\phi$ depends only on
one variable $\x$, $\sqrt{g}=V(\x)$, while the operator
is diagonal with respect to the angular modes $\exp[\i \om \phi]$.
The ratio \rf{calR} is then the product over angular
modes of the ratio of 1D determinants:
\be
\ln {\cal R}= \sum_\om 
 \ln\left[\frac{(-\partial^2+\om^2+\mu^2 V(\x))}{(-\partial^2+\om^2)}
\right].
\label{calRsum}
\ee
Here the sum over $\om$ runs over
integers or half-integers in bosonic or fermionic
determinants for the circular boundary, while $\om\in \Bbb R$
for its limiting case of the straight line, when
\be
\ln {\cal R}= \int \d\om
\ln \left[\frac{\det(-\partial^2+\om^2+\mu^2 V(\x))}{\det(-\partial^2+\om^2)}
\right].
\label{calRint}
\ee

Given the 1D determinants calculated above, it is possible to sum up
the angular modes in \eq{calRsum} or integrate in \eq{calRint}.
In particular, we can explicitly calculate the difference between
the sum and the integral to show that it indeed does
not depend on $\epsilon$ as was utilized in Sect.~\ref{s:res}.
This difference between the sum and the integral 
is computable by Plana's summation formula
\be
\frac 12 f(0)+\sum_{m=1}^\infty f(m)-\int_0^\infty \d \om\,f(\om)=
\i\int_0^\infty \d t\, \frac {f(\i t)-f(-\i t)}
{\e^{2\pi t}-1},
\label{Plan}
\ee
which holds when $f(z)$ is analytic for ${\rm Re}\; z\geq 0$, in particular,
on the imaginary axis. 

For the determinants in Eqs.~\rf{good}, \rf{67} with $\x_0=\eps$ we have
\be
f(m)=\ln (m+a),
\ee
and, using the formula (4.552) from Ref.~\cite{GR}, we obtain
for the right-hand side of \eq{Plan}
\be
-2\int_0^\infty \d t\, 
\frac{\arctan ( t/ a)}{\e^{2\pi t}-1}=
-\ln\Gamma(1+a)+\Big(a+\frac12\Big)\ln a-a  +\frac12 \ln(2\pi).
\ee
This results in the identity 
\bea
\lefteqn{
\frac 12 \ln a +\sum_{m=1}^\infty \ln(m+a) -\int _0^\infty \d\om\,\ln (\om+a)}
\non &=&
-\ln\Gamma(1+a)+\Big(a+\frac 12\Big)\ln a-a +\frac12 \ln(2\pi)
\stackrel{a\to\infty}\longrightarrow 
  {\cal O}(a^{-1}).
\label{smi}
\eea
Therefore, the integral exactly equals the sum as $a\sim 1/\eps\to\infty$
and no constant emerges in the difference.

An analog of \eq{smi} for $\x_0=0$ looks quite similar
when $a\ll 1/\eps \to \infty$
\bea
\lefteqn{\frac 12 {\rm Ei}\left( -2a\eps \right)+
\sum_{m=1}^\infty {\rm Ei}\left[ -2(m+a)\eps \right]-
\int_0^\infty \d\om\,{\rm Ei}\left[ -2(\om+a)\eps \right]}\non &&
\approx 
-\ln\Gamma(1+a)+
\Big(a+\frac 12\Big)\ln a-a +\frac12 \ln(2\pi),\hspace{0.7cm}
\mbox{for}\hspace{0.7cm}a\ll1/\eps.
\label{smi0}
\eea
This formula is derived in Appendix~\ref{appB}.
Again the difference between the sum and the
integral does not depend on $\eps$.

\subsection{Straight line}

It is also seen from the above formulas that
the difference between the sum and the
integral is UV finite. We should be careful, however, at this point because,
as is already pointed out in footnote$^{\ref{f:1}}$, 
the upper half-plane bounded by the straight
line with periodically identified ends has the Euler character zero
rather than one as for the disk. The logarithmic UV divergence
in the ratio of determinants is proportional to the (exponential of the)
difference of the Euler characters~\cite{Alv83}. This is demonstrated
by explicit calculations of this Subsection. Modulo this subtlety
both the dependence on a UV cutoff and
the $\epsilon$-dependence of the ratio of determinants for the circle
can be evaluated from the integral, associated with the ratio
of determinants for a straight line.
In the ratio of determinants~\rf{thefla} the UV divergence cancels,
since they are computed for the 
surfaces with the same Euler character.

Using the UV regularization $\Lambda$ by cutting off high frequencies
at $\om=\Lambda$, which
is described in Appendix~\ref{appR}, we obtain 
as $\Lambda\to\infty$ the following results. 

\paragraph{Massive bosons.} 
We get from \eq{ssb} for $\x_0=\eps$
\be
\int_0^\Lambda \d \om \,\ln\left(1+\frac1{\om\eps}  \right)
=\frac1\eps \Big(\ln(\Lambda \eps)+ 1 \Big),
\label{s2b}
\ee
and from \eq{ssb0} for $\x_0=0$
\be
\int_0^\Lambda \d \om \,\left[\frac{1-\e^{-2\om\eps}}{\om\eps}
-{\rm Ei}(-2\om\eps)\right]
=\frac1\eps \left(\ln(2\Lambda \eps)+\gamma_{\rm E}+\frac 12 \right).
\label{s1b}
\ee

\paragraph{Massive fermions.}
We get from \eq{ssf} for $\x_0=\eps$
\be
\int_0^\Lambda \d \om \,\ln\left(1+\frac1{2\om\eps}  \right)
=\frac1{2\eps} \Big(\ln(2\Lambda \eps)+ 1 \Big),
\label{s2f}
\ee
and from \eq{ssf0} for $\x_0=0$
\be
\int_0^\Lambda \d \om \,\left[\frac{1-\e^{-2\om\eps}}{2\om\eps}
-{\rm Ei}(-2\om\eps)+\frac 12 \e^{-2\om\eps} \right]
=\frac1{2\eps} \left(\ln(2\Lambda \eps)+\gamma_{\rm E}+\frac 32 \right).
\label{s1f}
\ee

\paragraph{Longitudinal modes.}
The logarithm of the right-hand side of \eq{ldf} for $\x_0=\eps$ equals
\be
\int_0^\Lambda \d \om\,\ln \left(1+\frac1{\eps\om}+\frac1{2 \eps^2 \om^2}\right)
= \frac1\eps\left( \ln(\Lambda \eps)+
1+\frac\pi4+\frac12\ln2 \right).
\ee

For $\x_0=0$ we analogously find
\be 
 \rf{ldf0}
= \frac1\eps \left( \ln(2\Lambda\eps)+\gamma_{\rm E} +\frac 74 \right).
\ee

\paragraph{Massless fermions.}
The logarithm of the right-hand side of \eq{mfd13} for $\x_0=\eps$ equals
\be
\int_0^\Lambda \d \om\,\ln \left(
 -2 \e^{2 \eps \om} \eps \om {\rm Ei}(-2 \eps \om) \right)=
\frac{1}{\eps}\left(-\frac12\ln(\Lambda\eps)-C_1   \right), \qquad
C_1=0.438934\,,
\label{ffee}
\ee
where the constant $C_1$ is found numerically. For the constant $C_2$
in \eq{C2} this gives
\be
C_2=\frac 32 \ln 2 -2 -\frac\pi4 +4 C_1 \approx 0.01 \,.
\label{CC2}
\ee
Because of the occurred cancellation we cannot exclude that $C_2$ is
actually zero. A more precise analysis is required at this point. 

For $\x_0=0$ the logarithm of \eq{mfd0} equals  
\be
\ln\;\rf{mfd0}= -\frac1{2\eps} 
\left(\ln(2\Lambda\eps)+\gamma_{\rm E}+1 \right).
\ee
For the constant $C_2$ in \eq{C2} this gives
\be
C_2=\frac74.
\ee

We see from the explicit computations of this Subsection that 
in the ratios of determinants the
$\frac 1\eps \ln \eps$ terms are
the same for $\x_0=\eps$ and $\x_0=0$, while the $1/\eps$ terms
change.

\pagebreak

\subsection{Circle}

\paragraph{Massive bosons.}
We obtain from \eq{good} for $\x_0=\eps$ 
\bea
\frac12\ln \frac1\eps
+\sum_{m=1}^\Lambda \left[ \ln \Big(m+\frac 1\eps\Big)-\ln(m+1)  \right]
= \left(\frac1\eps-1\right) \ln\Lambda +\frac1\eps \ln \eps
+\frac 1\eps-\frac 12 \ln(2\pi),~~
\label{c2b}
\eea
and from \eq{79} for $\x_0=0$
\bea
&&-\frac12 \Big[\ln(2\eps)+\gamma_{\rm E}-2\Big] +
\sum_{m=1}^\Lambda \frac 12 \left(\frac1\eps-1\right)
\Big\{ {\rm Ei}[-2(m-1)\eps]-  {\rm Ei}[-2(m+1)\eps] 
-\ln(m-1) \non &&+\ln(m+1)\Big\}
-\sum_{m=1}^\infty {\rm Ei}[-2(m+1)\eps]= 
 \left(\frac1\eps-1\right) \ln\Lambda +
\frac1\eps \left(\ln(2\eps)+\gamma_{\rm E}+\frac 12 \right)
-\frac 12 \ln(2\pi) . \non
\label{c1b}
\eea

\paragraph{Massive fermions.} 
Substituting $\om=m-1/2$, we get from \eq{67} for $\x_0=\eps$
\bea
\sum_{m=1}^\Lambda \left[ \ln \Big(m+\frac 1{2\eps}-\frac12\Big)-\ln m \right]
= \frac12\left(\frac1{\eps}-1\right) \ln\Lambda + \frac1{2\eps}
\ln (2\eps)+\frac 1{2\eps} -\frac 12 \ln(2\pi),
\label{c2f}
\eea
and from \eq{79f} for  $\x_0=0$
\bea
\sum_{m=1}^\Lambda \frac 12 \left(\frac1\eps-1\right)
\Big\{ {\rm Ei}[-2(m-1)\eps]-  {\rm Ei}[-2m\eps] 
-\ln(m-1)+\ln m\Big\}
-\sum_{m=1}^\infty {\rm Ei}[-2m\eps] \non
+\frac12 \sum_{m=1}^\infty \e^{-(2m-1)\eps}= 
\frac 12 \left(\frac1\eps-1\right) \ln\Lambda +
\frac1{2\eps} \left(\ln(2\eps)+\gamma_{\rm E}+\frac 32 \right) 
-\frac12\ln(2\pi).  ~~~~~~
\label{c1f}
\eea

\subsection{Circle minus straight line}

Subtracting the contributions of the circle and the straight line,
we obtain for the differences:
\begin{subequations}
\bea
\rf{c2b}-\rf{s2b}&= &-\ln \Lambda-\frac 12 \ln(2\pi),\\
\rf{c1b}-\rf{s1b}&= &-\ln \Lambda-\frac 12 \ln(2\pi),\\
\rf{c2f}-\rf{s2f}&= &-\frac 12 \ln \Lambda-\frac 12 \ln(2\pi),\\
\rf{c1f}-\rf{s1f}&= &-\frac12 \ln \Lambda-\frac 12 \ln(2\pi),
\eea
\end{subequations}
which coincide for $\x_0=\eps$ and  $\x_0=0$ both for bosons and 
fermions. In Appendix~\ref{appL} we reproduce this computation, using
an extension of the $\zeta$-function regularization.
 
The fact that the differences do not depend upon 
whether $\x_0=\eps$ or $\x_0=0$
seems to be quite nontrivial since the contributions 
of individual modes with a certain $m$ do change. 
Their product over $m$ gives, however, the same both for the
bosonic and fermionic determinants. The physical implication of
this fact is that only the distances far away from the boundary ($\gg \eps$)
contribute to the differences.

\section{Conclusion and Outlook\label{s:conclu}}

The main result of this Paper is based on an evaluation of
the ratio of 2D determinants, that emerge in the one-loop
effective action of the open Green--Schwarz superstring in
$AdS_5\times S^5$ background, for a circular boundary. 
We have concentrated on the dependence of the ratio on
the parameter $\eps$, regularizing the near-boundary singularity in
AdS, and have shown that it does not involve a term $\frac1\eps \ln \eps$
in the exponent,
coming from the bosonic and fermionic determinants, since this term cancels 
against the one coming from the longitudinal determinant.
We differ at this point from the results of Ref.~\cite{KT08},
where this term remained and a special procedure of dealing with it
was implemented. 
The only possible dependence of the ratio upon $\eps$ is like
the exponential of $1/\eps$ which is similar to that of
the classical action and is removable by a Legendre transformation.

The remaining $\eps$-dependence resides in the reparametrization 
path integral of the exponential of the classical boundary action
in AdS space,
which is explicitly constructed in Ref.~\cite{AM11}. It is
a counterpart of the one in flat space, 
known as Douglas' integral~\cite{Dou31},
whose minimization with respect to functions, reparametrizing  the
boundary, or equivalently boundary metrics, reconstructs a minimal surface.
The cancellation of determinants, mentioned in the previous paragraph, 
is like decoupling of the conformal factor (or the Liouville field) in
the bulk, which happens in the critical dimension $d=10$ for the
Green--Schwarz superstring.

If the cancellation of the bulk determinants continues to
higher loops, say because of supersymmetry, then the reparametrization
path integral itself may be equivalent to the exact result~\cite{ESZ00} for
the circle. This interesting possibility deserves further study.     

\begin{acknowledgments}
We are grateful to Jan Ambj\o rn, Alexander Gorsky, Andrey Marshakov,
Poul Olesen, Anton Zabrodin, Konstantin Zarembo for discussions
and to Martin Kruczenski, Alin Tirziu and Arkady Tseytlin
for email correspondence. C.K. was supported by FNU through grant number
272--08--0329.
Y.M. thanks the NBI High Energy Theory group for hospitality and 
financial support. 

\end{acknowledgments}

\appendix

\section{Semi-classical correction at large $\boldsymbol{\om}$ \label{appCC}}

We can compute the fermionic determinant at large $\om=r$ by a
semi-classical expansion in $1/\omega$. For $\om\gg 1/\eps$ 
we can find the diagonal resolvent by iteratively solving the
Gel'fand--Dikii equation~\rf{GDe}:
\be
R_\om (\x,\x;V)=\frac 1{2\om}-\frac {V_{f\pm}(\x)}{4 \om^3}+{\cal O}(\om^{-5}),
\label{1stGD}
\ee
where $V_{f\pm}$ is given by \eq{Vf} with $\om$ substituted by $r$.
The meaning of this procedure is that we include this factor in the
definition of the potential, performing the expansion
in the inverse spectral parameter $\om$ for an arbitrary potential. 

The $1/\om^3$ term in \eq{1stGD}
coincides with the first Gel'fand--Dikii differential polynomial.
Integrating over $\x$ and $\om$, we obtain for the logarithm of the ratio 
\be
\int_r^\infty \d \om \frac 1{2\om^2 } \int_\eps^\infty \d \x\, V_{f\pm}(\x)
=\pm \frac 12 \int_\eps^\infty \d \x\, (\coth\x-1)=
\frac12\left[\eps -\ln(2\sinh\eps)\right]\stackrel{\eps\to0}\longrightarrow 
\mp\frac12\ln(2\eps).
\ee
It does not vanish as $r\to\infty$ because $V_{f\pm}\propto r$ itself.

We are now in a position to determine the constant $C$ introduced in \eq{77f+}.
Noting that  $\ln {\cal R}_{f+}\to C$ from \eq{77f+}, we deduce that
\be
C=\frac 12 \ln \frac{\coth\eps+1}2\stackrel{\eps\to0}
\longrightarrow-\frac12\ln(2\eps).
\label{C=}
\ee 
This reproduces the result of Ref.~\cite{KT08}.

\section{Summing over angular modes\label{appB} for 
$\boldsymbol{\x_0\!=\!0}$}

The sum over angular modes, i.e.\ over $m$ in the bosonic case or
$r=m+1/2$ in the fermionic case, involves  for $\x_0=0$ an exponential of
the sum
\be
\sum_{m=1}^\infty {\rm Ei}\left[-2(m+a)\eps  \right]=
- \int_{2\eps}^\infty \frac{\d t}t \frac{\e^{-a t}}{(\e^t-1)},
\label{thesum}
\ee
which is obviously convergent for $a>-1$. 
Notice that this produces something different 
from what results from first expanding in $\eps$ and
then summing over $m$. The summation is hence not interchangeable with
the $\eps\to0$ limit.

For small $\eps$ and $a\sim1$ 
we can evaluate the integral in \eq{thesum} as
\be
\int_{2\eps}^\infty \frac{\d t}t \frac{\e^{-a t}}{(\e^t-1)}
=\frac1{2\eps} + \left(a+ {\frac12}\right)
\left[\, \ln (2\eps) + \gamma_{\rm E}\right]+\ln\Gamma(1+a)-\frac 12 \ln(2\pi)
-\sum_{n=2}^\infty B_n(-a) \frac{(2\eps)^{n-1}}{(n-1)n!},
\label{B2}
\ee
where
\be
B_0(-a)=1,\qquad B_1(-a)=-a-\frac12, \qquad B_2(-a)=a^2+a+\frac16,
\ee
are the Bernoulli polynomials.

In \eq{B2} $1/\eps$ remarkably emerges as a regularized contribution of 
high modes. If we substituted each term in the sum \rf{thesum}
by its $\eps\to0$ asymptote and used the $\zeta$-function regularization,
we would not get this term. This happens because the $m\to \infty$
and $\eps\to0$ limits do not commute.

The integral which is the companion of the sum \rf{thesum} reads
\be
-\int_0^\infty \d \om\,{\rm Ei}\left[ -2(\om+a)\eps \right]= 
\int_{2\eps}^\infty \frac{\d t}{t^2}\e^{-a t} \approx 
\frac{1}{2\eps}+a \left[\ln(2a\eps) +\gamma_{\rm E}-1\right]-a^2\eps
+{\cal O}(\eps^2),
\ee
for $a\ll1/\eps$.
We can calculate them separately because
both the sum and the integral are convergent.
The difference of the two reproduces \eq{smi0}.

\section{Regularization by ``proper frequency''\label{appR}}

The usual proper time regularization of functional determinants is 
defined as
\be
-\tr \ln (-\Delta+V)\Big|_{\rm reg}= \int \d x \int^\infty_{\Lambda^{-2}}
\frac{\d t}t \,\LA x|\e^{t(\Delta-V)} |x\RA =
 \int \d x \int^\infty_{1}
\frac{\d \tau}\tau \,\LA x|\e^{\tau(\Delta-V)/\Lambda^2} |x\RA .
\ee
We use instead the regularization consisting in
cutting off very high frequencies:
\bea
-\tr \ln (-\Delta+V)\Big|_{\rm reg}&= &\int \d x 
\int_0^{\Lambda^{2}} \d \om^2\, R_\om(x,x;V)\non &=&
\int \d x 
\int_0^{\Lambda^{2}} \d \om^2\, \int_0^\infty
\d t \,\LA x|\e^{t(\Delta-\om^2-V)} |x\RA \nonumber \\ &= &
 \int \d x \int^\infty_{0}
\frac{\d t}t \,\left(1-\e^{-t \Lambda^2} \right)
\LA x|\e^{t(\Delta-V)} |x\RA \non &= &
 \int \d x \int^\infty_{0}
\frac{\d \tau}\tau \,\left(1-\e^{-\tau} \right)
\LA x|\e^{\tau(\Delta-V)/\Lambda^2} |x\RA .
\eea
The two look similar, the difference shows up only
for eigenvalues of the order of the cutoff.

\section{Summing using the Lerch function\label{appL}}

An alternative to the regularization by cutting off high frequencies 
from Appendix~\ref{appR} 
is the ``supersymmetric summation'' used in \cite{KT08}, where
the following series appears
\be
\sum_{m=1}^\infty \e^{-\mu m}\Big[ \ln(m+a)-\ln(m+b) \Big]
=-
\int_0^\infty \frac{\d t}t 
\frac{(\e^{-a t}-\e^{-b t})}{\e^{t+\mu}-1}.
\label{resum}
\ee
We have used here the integral representation of the Lerch transcendent
\be
\sum_{m=1}^\infty \frac{z^m}{(m+a)^s}
=\frac{z}{\Gamma(s)} \int_0^\infty \d t\, 
\frac{t^{s-1}\e^{-a t}}{\e^t-z}.
\label{Lerch}
\ee
To prove \eq{resum}, we expand
\be
\frac{1}{\e^{t+\mu}-1}=\sum_{m=1}^\infty \e^{-m(t+\mu)},
\ee
and use
\be
-\int_0^\infty \frac{\d t}t \left[\e^{-(a+m)t}- \e^{-(b+m)t}\right]=
\ln(m+a)-\ln(m+b). 
\ee

To compute the integral in \eq{resum}, we rewrite it as
\bea
-\int_0^\infty \frac{\d t}t 
\frac{(\e^{-a t}-\e^{-b t})}{(\e^{t+\mu}-1)}=
-\int_0^\infty \frac{\d t}t 
(\e^{-a t}-\e^{-b t})\left(\frac1{\e^{t+\mu}-1}-\frac1{t+\mu}+\frac1{t+\mu}
\right).~~~
\eea
We have
\bea
-\int_0^\infty \frac{\d t}t 
\frac{(\e^{-a t}-\e^{-b t})}{t+\mu}&=&
\frac1\mu \ln\frac ab-\frac1\mu\left[\e^{a\mu} {\rm Ei}(a\mu)- 
\e^{b\mu} {\rm Ei}(b\mu)  \right]\non
&=&
\left[(a-b)\left(\ln \frac1\mu -\gamma_{\rm E}+1 \right)
- a \ln a + b \ln b \right]+{\cal O}(\mu),
\eea
and
\be
-\int_0^\infty \frac{\d t}t 
(\e^{-a t}-\e^{-b t})\left(\frac1{\e^{t}-1}-\frac1{t}\right)
=-\ln \Gamma(1+a)+ a\ln a-a
 +\ln \Gamma(1+b)- b\ln b-b.
\ee
For small $\mu$ we obtain for the integral on the right-hand side of 
\eq{resum}:
\be
{\rm \rf{resum}}=(a-b)\left(\ln \frac1\mu -\gamma_{\rm E} \right)-\ln\Gamma(1+a)
 +\ln \Gamma(1+b)+{\cal O}(\mu),
\ee
which coincides with the formula 
\be
\sum_{m=1}^\Lambda \left[ \ln(m+a)-\ln(m+b)  \right]=(a-b)\ln\Lambda-
\ln\Gamma(1+a)+\ln\Gamma(1+b),
\label{sumlogs}
\ee
obtained by using \eq{smi} with the regularization of Appendix~\ref{appR},
if
\be
\Lambda=\frac 1\mu \e^{-\gamma_{\rm E}}.
\ee

For the bosonic determinants we obtain:
circle for $\x_0=\eps$
\be
\frac 12 \ln \frac 1\eps+
\sum_{m=1}^\infty \e^{-\mu m} \left[\ln \Big(m+\frac 1\eps\Big)-\ln(m+1)\right]
= \left(\frac1\eps-1\right) \ln\Lambda +\frac1\eps\ln \eps
+\frac 1\eps-\frac 12 \ln(2\pi),
\label{c2baa}
\ee
and circle for  $\x_0=0$
\bea
-\frac 12\left[ \ln(2\eps)+ \gamma_{\rm E}-2\right]+
\sum_{m=1}^\infty \frac 12 \left(\frac1\eps-1\right)
\Big\{ {\rm Ei}[-2(m-1)\eps]-  {\rm Ei}[-2(m+1)\eps]  \non 
+ \e^{-\mu m} \left[-\ln(m-1)+\ln(m+1) \right]\Big\} 
-\sum_{m=1}^\infty {\rm Ei}[-2(m+1)\eps]\non= 
 \left(\frac1\eps-1\right) \ln\Lambda +
\frac1\eps \left(\ln(2\eps)+\gamma_{\rm E}+\frac 12 \right)
-\frac12\ln(2\pi),\non
\label{c1baa}
\eea
straight line for $\x_0=\eps$
\be
\int_0^\infty \d \om \,
\e^{-\mu\om}\ln\left(1+\frac1{\om\eps}  \right)
=\frac1\eps \Big(\ln(\Lambda \eps)+ 1 \Big)
\label{s2baa},
\ee
and straight line for  $\x_0=0$ 
\be
\int_0^\infty \d \om \,\e^{-\mu\om}\left[\frac{1-\e^{-2\om\eps}}{\om\eps}
-{\rm Ei}(-2\om\eps)\right]
=\frac1\eps \left(\ln(2\Lambda \eps)+\gamma_{\rm E}+\frac 12 \right).
\label{s1baa}
\ee
For the difference of the circle and the straight line we find
\begin{subequations}
\bea
{\rm \rf{c2baa}}-{\rm \rf{s2baa}}=-\ln \Lambda -\frac 12 \ln(2\pi),
\label{sbaapp} \\*
{\rm \rf{c1baa}}-{\rm \rf{s1baa}}=-\ln \Lambda -\frac 12 \ln(2\pi),
\label{sbaap}
\eea
\end{subequations}
which again coincide. The divergent term in both cases is owing to
the difference in the Euler characters (see footnote$^{\ref{f:1}}$).

Fermions: circle for $\x_0=0$ with positive $r=m-1/2$ ($m\geq 1$)
\bea
\sum_{m=1}^\infty \e^{-\mu r}\ln {\cal R}_{f+}\Big|_{\x_0=0} &=&
\sum_{m=1}^\infty  \e^{-\mu (m-1/2)}
\frac14 \left( \coth\eps-1 \right)
\Big\{ {\rm Ei} \left[-2(m-1)\eps\right] -
{\rm Ei} \left[-2m\eps\right]  \non  &&
\hspace*{1cm}
-\ln(m-1)+\ln m
\Big\} +
\sum_{m=1}^\infty  \e^{-\mu(m-1/2)}C \nonumber \\
 &=& \frac14 \left( \coth\eps-1 \right)\left[
\ln(2\eps)+\gamma_{\rm E}+\ln \frac 1\mu \right]
+C \frac 1\mu .
\label{77f+0p}
\eea
where $C$ is given by \eq{C=}, and with negative $r=-m-1/2$ ($m\geq 0$)
\bea
\sum_{m=0}^\infty \e^{-\mu |r|}\ln {\cal R}_{f-}\Big|_ {\x_0=0}
&=&
\sum_{m=0}^\infty  \e^{-\mu (m+1/2)}\left\{
\frac{1}{4} (\coth\eps-1) \Big[ \text{Ei}(-2m\eps)
-\ln m +\ln (m+1) \Big] \right.\non 
&& 
\left.- \frac 14\left(\coth\eps+3 \right) \text{Ei}[-2(m+1)\eps]
+\frac {1}{2\eps} \e^{-(2m+1) \eps} -\,C \right\} \nonumber \\
&=& \frac14 \left( \coth\eps-1 \right)\left[
\ln(2\eps)+\gamma_{\rm E}+\ln \frac 1\mu \right]+\frac{3}{4\eps}
+\frac12 \left[\ln(2\eps)+\gamma_{\rm E} \right] \non
&& - \frac 12 \ln (2\pi)
-C \frac 1\mu .
\label{77f-0p}
\eea
For the sum we find
\be
{\rm \rf{77f+0p}}+
{\rm \rf{77f-0p}}=\frac12 \left( \coth\eps-1 \right)\ln \Lambda
+\frac1{2\eps}\left[
\ln(2\eps)+\gamma_{\rm E}+\frac 32 \right]
 - \frac 12 \ln (2\pi) . 
\label{fsumaa}
\ee

The contribution of the straight line ($\x_0=0$) is
\be
\int_0^\infty \d \om \,\e^{-\mu \om}\left[\frac{1-\e^{-2\om\eps}}{2\om\eps}
-{\rm Ei}(-2\om\eps)+\frac 12 \e^{-2\om\eps} \right]
=\frac1{2\eps} \left[\ln\Lambda+
\ln(2\eps)+\gamma_{\rm E}+\frac 32 \right],
\label{s1faa}
\ee
so that we find for the difference 
\be
{\rm \rf{fsumaa}}-{\rm \rf{s1faa}}=-\frac12\ln \Lambda - \frac 12 \ln(2\pi).
\label{D21}
\ee

Fermions with an alternative ``bosonic'' regularization:
circle for $\x_0=0$ with positive $r=m-1/2$ ($m\geq 1$)
\bea
\sum_{m=1}^\infty \e^{-\mu |r+1/2|}\ln {\cal R}_{f+}\Big|_{\x_0=0}
&=&
\sum_{m=1}^\infty  \e^{-\mu m}
\frac14 \left( \coth\eps-1 \right)
\Big\{ {\rm Ei} \left[-2(m-1)\eps\right] -
{\rm Ei} \left[-2m\eps\right]  \non  &&
\hspace*{1cm}
-\ln(m-1)+\ln m
\Big\} +
\sum_{m=1}^\infty  \e^{-\mu(m-1/2)}C \nonumber \\
 &=& \frac14 \left( \coth\eps-1 \right)\left[
\ln(2\eps)+\gamma_{\rm E}+\ln \frac 1\mu \right]
+C \left[\frac 1\mu -\frac 12 \right].
\label{77f+0pal}
\eea
and with negative $r=-m-1/2$ ($m\geq 0$)
\bea
\sum_{m=0}^\infty \e^{-\mu |r+1/2|}\ln {\cal R}_{f-}\Big|_{\x_0=0}
 &=&
\sum_{m=0}^\infty  \e^{-\mu m}\left\{
\frac{1}{4} (\coth\eps-1) \Big[ \text{Ei}(-2m\eps)
-\ln m +\ln (m+1) \Big] \right.\non 
&& 
\left.- \frac 14\left(\coth\eps+3 \right) \text{Ei}[-2(m+1)\eps]
+\frac {1}{2\eps} \e^{-(2m+1) \eps} -\,C \right\} \nonumber \\
&=& \frac14 \left( \coth\eps-1 \right)\left[
\ln(2\eps)+\gamma_{\rm E}+\ln \frac 1\mu \right]+\frac{3}{4\eps}
+\frac12 \left[\ln(2\eps)+\gamma_{\rm E} \right] \non
&& - \frac 12 \ln (2\pi)
-C \left[\frac 1\mu +\frac 12 \right].
\label{77f-0pal}
\eea
For the sum we find
\be
{\rm \rf{77f+0p}}+{\rm \rf{77f-0p}}=
\frac12 \left( \coth\eps-1 \right)\ln \Lambda
+\frac1{2\eps}\left[ \ln(2\eps)+\gamma_{\rm E}+\frac 32 \right]
 - \frac 12 \ln (2\pi) -C, 
\label{fsumaaal}
\ee
and
\be
{\rm \rf{fsumaaal}}-
{\rm \rf{s1faa}}=-\frac 12 \ln \Lambda - \frac 12 \ln(2\pi) -C,
\ee
which differs from \rf{D21}.

We have thus reproduced in this Appendix the results of Sect.~\ref{s:sum}.


\end{document}